# Uniqueness of conserved currents in quantum mechanics

**Peter Holland**

Green College
University of Oxford
Woodstock Road
Oxford OX2 6HG
England

Email: peter.holland@green.ox.ac.uk

**Abstract**

It is proved by a functional method that the conventional expression for the Dirac current is the only conserved 4-vector implied by the Dirac equation that is a function of just the quantum state. The demonstration is extended to derive the unique conserved currents implied by the coupled Maxwell-Dirac equations and the Klein-Gordon equation. The uniqueness of the usual Pauli and Schrödinger currents follows by regarding these as the non-relativistic limits of the Dirac and Klein-Gordon currents, respectively. The existence and properties of further conserved vectors that are not functions of just the state is examined.

# 1 Introduction

The derivation of an expression for the conserved current density associated with a particular wave equation is a standard textbook demonstration in quantum mechanics. The current (we use the term here to include four components, a density $\rho$ and a 3-vector density **j**), of course, plays a central role in the physical applications of quantum theory. For example, it encodes the probability distributions of quantum mechanics, and in QED the Dirac current is a source in Maxwell's equations. Yet, in a strange oversight, the issue of the uniqueness of the usual quoted expressions for current does not appear to have been addressed in the standard sources.

The question raised here is whether the usual expressions really are 'unique'. By this we mean that the functional dependence of the current on the wavefunction is uniquely determined by the associated wave equation. That the conventional currents are the correct ones is surely not in doubt, but how might this be proved, and under what assumptions? To illustrate what a possible nonuniqueness of the current would mean, consider the conservation equation implied by the Schrödinger equation:

$$\frac{\partial \rho}{\partial t} + \nabla \cdot \mathbf{j} = 0. \tag{1.1}$$

The usual expressions for the functions appearing in (1.1) are (for a scalar external potential)

$$\rho = \Psi^*\Psi, \quad \mathbf{j} = (\hbar/2mi)\left(\Psi^*\nabla\Psi - (\nabla\Psi^*)\Psi\right) \tag{1.2}$$

where $\Psi$ is the wavefunction. The possibility of a nonuniqueness in the expressions (1.2) arises because there is a 'gauge' freedom in (1.1): any scalar $\bar{\rho}$ and vector $\bar{\mathbf{j}}$ that differ from the traditional choices $\rho$ and **j** by additive fields $a$ and **a** (which may depend on $\Psi$), respectively, will be compatible with the Schrödinger equation if the additional fields also obey a conservation equation:

$$\bar{\rho} = \rho + a, \quad \bar{\mathbf{j}} = \mathbf{j} + \mathbf{a}, \quad \frac{\partial a}{\partial t} + \nabla \cdot \mathbf{a} = 0. \tag{1.3}$$

What arguments can we use to fix the additional variables? On empirical grounds one might argue that the scalar density is well established to be $\rho$ (so that $a = 0$). The vector **a** could be fixed by appealing to situations where the current itself (rather than its derivatives) is relevant. This happens when the quantum system interacts with other systems, and it appears that such a case can indeed be used to determine the current uniquely. Thus, in the analogous problem for a relativistic charged quantum particle mutually interacting with an electromagnetic field, it may be argued that the requirements that the Lagrangian be Lorentz and gauge invariant uniquely fix the form of the 4-current that appears in the interaction term [1]. This is in accordance with the principle of minimal coupling. However, this argument will not work in the case where the quantum system is subject only to external forces, and it appears to make the form of a quantity that one might expect is intrinsic to the system depend on the details of couplings with further systems. We note that Noether's theorem, which provides a formula to calculate the local conserved current associated with the gauge symmetry of the Lagrangian, is of no help in this regard since the formula does not imply a unique expression. This is because the Lagrangian is not unique; a given set of field equations will be compatible with Lagrangian densities which differ by a divergence, or which differ even more generally.

In the case of a spin $\frac{1}{2}$ particle subject only to external forces, a solution to this problem has been found by going to relativity theory [2]. The argument, which demonstrates the uniqueness of the usual expression for the Dirac current, follows from the combination of Lorentz covariance and the probability interpretation of the wavefunction (in a suitably low-energy regime where this is meaningful). For a spin $\frac{1}{2}$ particle whose associated spinor field $\psi^a(x^i,t)$ obeys the Dirac equation (see (2.1)), the Dirac probability current satisfies the conservation law

$$\partial_\mu J^\mu = 0, \quad J^\mu = c\overline{\psi}\gamma^\mu\psi, \quad \overline{\psi} = \psi^+\gamma^0. \tag{1.4}$$

Consider a new current $j^\mu$ that differs from $J^\mu$ by the addition of a divergenceless 4-vector $a^\mu$:

$$j^\mu = J^\mu + a^\mu, \quad \partial_\mu j^\mu = 0. \tag{1.5}$$

The new current is conserved and so could potentially be used to define a probability 4-vector. In order that the new current still reproduces the quantal distribution $J^0/c = \psi^+\psi$ we must have $a^0 = 0$. Now consider the current in another Lorentz frame, with spinor $\psi'(x'^i,t')$. Then in this frame $J'^0 = c\psi'^+\psi'$ and again $a'^0 = 0$. But the only 4-vector whose zeroth component vanishes in all frames is the zero vector. Hence $a^\mu = 0$ and the Dirac current is unique (the demonstration can be extended to the many-particle case [2]). Returning to the non-relativistic problem, this argument can be used to fix the functions $a$ and **a** by treating (1.1) and (1.3) as non-relativistic limits of the relativistic equations (1.4) and (1.5). As we shall see later (§5), the form of **a** depends on the system described by the Schrödinger equation.

A drawback with this proof is that it assumes that the correct expression for the zeroth component of the relativistic current has already been identified. Although, as mentioned above, this may be justified *a posteriori* by comparison with experiment, it seems reasonable to ask whether an *ab initio* proof of the uniqueness of the Dirac current (and other currents) that proceeds without first making this assumption can be found. The aim of this paper is to analyse to what extent we may answer this question by adapting a method employed by Fock [3] who investigated the analogous problem of proving the uniqueness of relativistic energy-momentum tensors. As with quantum-mechanical currents, this is important as these tensors themselves have physical significance (rather than just their integrals or derivatives; e.g., they appear as the source in Einstein's gravitational field equations). Fock was concerned with deriving a unique form for an energy-momentum tensor (up to additive and multiplicative constants) by assuming only the following: (a) the field equations obeyed by the relevant physical system, (b) the transformation properties of the energy-momentum tensor under Lorentz transformations, (c) the conservation law for this tensor, and (d) that the tensor depends only on the functions of state. His key step is to express the conservation equation as a functional relation for the fields, from which an identity can be extracted. The latter implies a set of equations for the tensor components whose solution leads to a complete solution of the problem. We shall proceed in the same way here, except of course application to a conserved 4-vector associated with a quantum-mechanical wave equation necessitates certain modifications to Fock's method.

It turns out that the analogues of Fock's assumptions, that the current is a conserved Lorentz 4-vector that is implied by Dirac's equation and that depends just on the quantum state, are sufficient to fix uniquely the form of the current as a function of the state. It is the last assumption, the analogue of Fock's requirement (d) (which Fock regarded as a new physical principle), that enforces uniqueness. However, this only partially answers our question since, as we shall see, there exist other conserved 4-vectors associated with the Dirac equation. A benefit of the theorem is that it establishes something about the character of the other vectors - for example, that they depend on functions other than the state variables (derivatives of the latter, or other arbitrary functions). Since the further vectors we consider generate the same global conserved quantity, and apparently are otherwise acceptable, the problem reduces to justifying why the 4-vector that depends only on the state functions has preferential status.

In §2 we give an argument to justify the assumption that the current is a function of just the state variables, and on this basis show that, up to additive and multiplicative constants, the only conserved 4-vector implied by the Dirac equation is the conventional expression (1.4). Under the same assumption of functional dependence, the method is extended to derive the unique particle currents associated with the coupled Maxwell-Dirac equations (§3) and the spin 0 Klein-Gordon equation (§4). Regarding the non-relativistic Pauli and Schrödinger currents as limits of the Dirac and Klein-Gordon currents, respectively, then establishes their uniqueness too (§5). The existence and properties of further conserved vectors that are not functions of just the state is examined in §2.4 and §4.2, and the status of Fock's assumption (d) is assessed in §6. The discussion is confined to single-particle, first-quantized quantum theory; the extension to many bodies and field theories will be considered elsewhere.

The uniqueness problem has been discussed previously [2,4,5] in the context of the de Broglie-Bohm pilot-wave theory [6-8]. This posits an equation of motion for a particle that is derived from the quantum-mechanical current so our results are of importance to that theory, as will be discussed elsewhere. Our demonstration, although relevant to the pilot-wave interpretation, is independent of that particular application.

*Notation*: Repeated indices are summed over and have the following values: Greek $\mu, \nu = 0,1,2,3$; Latin $a,b,c = 1,2,3,4$; $i,j,k = 1,2,3$; $A = 1,2$; $Q=1,\ldots,16$.

## 2 The Dirac current

### 2.1 Functional dependence of the current

The Dirac equation for a particle of mass *m* and charge *e* placed in an external electromagnetic field $A_\mu$ is [9]

$$\gamma^\mu \left( i\hbar \partial_\mu - e A_\mu \right) \psi = mc\psi. \tag{2.1}$$

The adjoint equation is

$$\overline{\psi} \left( -i\hbar \overleftarrow{\partial}_\mu - e A_\mu \right) \gamma^\mu = mc\overline{\psi}. \tag{2.2}$$

Here $\psi = \psi(x)$, $\psi = (\psi^a)$, $\psi^+ = (\psi^{*a})$, $\bar{\psi} = \psi^+ \gamma^0$, $x = (x^i, ct)$, $\partial_\mu = \partial/\partial x^\mu$, $c$ is the speed of light, and the $\gamma$ matrices obey the relations

$$\gamma^\mu \gamma^\nu + \gamma^\nu \gamma^\mu = 2\eta^{\mu\nu}, \quad \eta^{\mu\nu} = (1, -1, -1, -1). \tag{2.3}$$

We have suppressed the spinor indices but later it will be important to insert them. The following representation for the $\gamma$ matrices is used:

$$\gamma^0 = \begin{pmatrix} I & 0 \\ 0 & -I \end{pmatrix}, \quad \gamma^i = \begin{pmatrix} 0 & \sigma^i \\ -\sigma^i & 0 \end{pmatrix}, \quad \gamma^5 = i\gamma^0 \gamma^1 \gamma^2 \gamma^3 = \begin{pmatrix} 0 & I \\ I & 0 \end{pmatrix} \tag{2.4}$$

where $\sigma^i$, $i = 1, 2, 3$, are the Pauli matrices and I is the 2x2 unit matrix.

We assume that these equations imply an associated Lorentz 4-vector $j^\mu$ that obeys the conservation equation:

$$\partial_\mu j^\mu = 0 \tag{2.5}$$

(we know one exists – pre-multiplying (2.1) by $\bar{\psi}$ and post-multiplying (2.2) by $\psi$ and subtracting gives $J^\mu$ in (1.4)). Our aim is to use the set of equations (2.1), (2.2) and (2.5) to fix the functional dependence of $j^\mu$ on $\psi$ and $\psi^+$ (and possibly other fields). Subject to the considerations of the following paragraphs, it turns out that we need make no further assumptions about $j^\mu$ (e.g., that it is a proper vector, or real) to arrive at a unique expression.

To proceed, we must establish the general functional dependence of the current on the fields. First, since the wave equation is purely local, it is natural to assume a local dependence on $\psi$ and $\psi^+$, the latter being regarded as independent variables (equivalent to 8 independent real variables). These functions (for all $x^\mu$) entirely characterize the state of the system, since, given their initial values, the wave equation determines them for all time, and from them we can compute all other relevant functions. They are the 'functions of state'. What about dependence of the conserved vector on the derivatives of the fields? The following argument justifies a dependence only on the state variables (this is explored further in §2.4 and §6).

We recall that by a 'conserved quantity' $P$ in physics we mean a quantity that is conserved in time by virtue of the temporal evolution determined by the dynamical equations of the theory. This quantity therefore results, in effect, from integrating the dynamical equations with respect to the time. It follows that $P$ will depend on time derivatives of the state variables of the theory of order one less than appear in the dynamical equations. Thus, we might expect that $P$ is a function only of the state variables. For example, in particle mechanics the dynamical equation involves $\ddot{x}$ and the conserved quantities - energy, momentum and angular momentum - are functions of the variables $x$ and $\dot{x}$. In this example the conserved quantities are indeed functions of just the state variables but in the case of a field theory the situation is more complicated. In that case the quantity $P$ will either be automatically free of the time derivatives of the fields, or it can be made so by substituting for the time derivatives from the field equation. However, in both cases the space derivatives of the fields may appear. Then it may not be the case that $P$ depends solely on the state variables, if the latter do not include all the derivatives that appear (for examples see §2.4 and §6). Thus, unlike the particle case, for fields it may not always be possible to achieve dependence of $P$ on just the state, and the possibility of doing this must be checked on a case-by-case basis.

In the present case, where

$$P = \int j^0 d^3x, \quad \frac{dP}{dt} = 0, \tag{2.6}$$

the dynamical equation (2.1) is first order in time. Hence, pursuing our argument, we require that $P$ does not depend on $\partial \psi / \partial t$ or $\partial \psi^+ / \partial t$. This can be achieved if $j^0$ does not depend on these quantities. We now make the assumption that $j^\mu$ is a covariant function of its arguments. Then the components $j^i$ will likewise not depend on the time derivatives, and the spatial derivatives of $\psi$ and $\psi^+$ will also be absent from $j^\mu$. We conclude that the functional dependence of $j^\mu$ on the fields is purely through the state variables, $\psi$ and $\psi^+$.

The only other possible dependence of $j^\mu$ is on fields that are independent of $\psi$ and $\psi^+$, including explicit functions of the coordinates. We should include the possibility that the potentials $A_\mu$, for example, could appear in the current since they are present in (2.1). We shall, however, exclude further arbitrary functions since such dependence would imply that the current is not a property purely of the wave equation. We thus have finally that $j^\mu = j^\mu(\psi^a, \psi^{*a}, A_\nu)$. It is important to note though that, whatever the merits of this argument, its implementation does not produce the *only* conserved vector associated with the Dirac equation, as we shall see in §2.4.

## 2.2 Derivation of the unique current

Following Fock [3], the method proceeds by recasting equation (2.5) as a constraint on the functional dependence of the current on the fields, using the usual 'function of a function' formula:

$$\frac{\partial j^\mu}{\partial \psi^a} \partial_\mu \psi^a + \frac{\partial j^\mu}{\partial \psi^{*a}} \partial_\mu \psi^{*a} + \frac{\partial j^\mu}{\partial A_\nu} \partial_\mu A_\nu = 0. \tag{2.7}$$

This is the basic relation from which we shall derive an expression for $j^\mu$.

Our first observation is that the spacetime derivatives of $A_\mu$ are absent from the first two members in (2.7), by virtue of the field equations (2.1) and (2.2). The final term in (2.7) must therefore vanish. Since the derivatives of $A_\mu$ are arbitrary, we conclude that $\partial j^\mu / \partial A_\nu = 0$, i.e., $j^\mu$ is independent of $A_\mu$. Thus, the only dependence of $j^\mu$ on the spacetime coordinates is through the $\psi$-field, and our basic relation reduces to

$$\frac{\partial j^\mu}{\partial \psi^a} \partial_\mu \psi^a + \frac{\partial j^\mu}{\partial \psi^{*a}} \partial_\mu \psi^{*a} = 0 \tag{2.8}$$

or in matrix terms

$$\frac{\partial j^\mu}{\partial \psi} \partial_\mu \psi + \partial_\mu \psi^+ \frac{\partial j^\mu}{\partial \psi^+} = 0. \tag{2.9}$$

The principal component of our demonstration is to derive from (2.8) an expression for $j^0$. The components $j^i$ will then be seen to follow straightforwardly.

In (2.8) the field and its derivatives are connected by the wave equation (2.1), and the conjugate field and its derivatives by the adjoint equation (2.2). To remove this connection between the fields, and thereby obtain an identity involving the independent quantities that remain, we replace the time derivatives of the fields appearing in (2.8) by substituting from the field equations. The field equations (2.1) and (2.2) imply

$$i\hbar \partial_0 \psi = -i\hbar \gamma^0 \gamma^i \partial_i \psi + e\gamma^0 \gamma^\mu A_\mu \psi + mc\gamma^0 \psi \\ i\hbar \partial_0 \psi^+ = -i\hbar \partial_i \psi^+ \gamma^0 \gamma^i - e\psi^+ \gamma^0 \gamma^\mu A_\mu - mc\psi^+ \gamma^0 \qquad (2.10)$$

and hence

$$(i\hbar)^{-1} \frac{\partial j^0}{\partial \psi}\left(e\gamma^0 \gamma^\mu A_\mu \psi + mc\gamma^0 \psi\right) - \frac{\partial j^0}{\partial \psi}\gamma^0 \gamma^i \partial_i \psi + \frac{\partial j^i}{\partial \psi}\partial_i \psi \\ -(i\hbar)^{-1}\left(e\psi^+ \gamma^0 \gamma^\mu A_\mu + mc\psi^+ \gamma^0\right)\frac{\partial j^0}{\partial \psi^+} - \partial_i \psi^+ \gamma^0 \gamma^i \frac{\partial j^0}{\partial \psi^+} + \partial_i \psi^+ \frac{\partial j^i}{\partial \psi^+} = 0, \qquad (2.11)$$

This relation is an identity in which $\psi$, $\psi^+$, $\partial_i \psi$ and $\partial_i \psi^+$ are independent variables. Since $j^\mu$ depends on the fields but not on their derivatives, the coefficients of $\partial_i \psi$ and $\partial_i \psi^+$ for each $i$ must vanish. In addition, $A_\mu$ is an arbitrary function so its coefficient vanishes. We therefore deduce the following four relations:

$$\frac{\partial j^0}{\partial \psi}\gamma^0 \psi - \psi^+ \gamma^0 \frac{\partial j^0}{\partial \psi^+} = 0 \qquad (2.12)$$

$$\frac{\partial j^0}{\partial \psi}\gamma^0 \gamma^\mu \psi - \psi^+ \gamma^0 \gamma^\mu \frac{\partial j^0}{\partial \psi^+} = 0 \qquad (2.13)$$

$$-\frac{\partial j^0}{\partial \psi^a}\left(\gamma^0 \gamma^i\right)^a_b + \frac{\partial j^i}{\partial \psi^b} = 0 \qquad (2.14)$$

$$-\left(\gamma^0 \gamma^i\right)^c_a \frac{\partial j^0}{\partial \psi^{*a}} + \frac{\partial j^i}{\partial \psi^{*c}} = 0. \qquad (2.15)$$

To determine from these relations the functional dependence of $j^0$, we shall use two of its transformation properties. The first is that $j^0$ is a scalar with respect to a spatial rotation. Hence, $j^0$ is a function only of quantities that are scalars under spatial rotations. There are four such quantities that one can construct from the bilinear covariants implied by a Dirac spinor,

$$\beta = \overline{\psi}\psi, \; V^\mu = \overline{\psi}\gamma^\mu \psi, \; S^\mu = \overline{\psi}\gamma^\mu \gamma^5 \psi, \; B^{\mu\nu} = i\overline{\psi}\gamma^{\mu\nu}\psi, \; \chi = i\overline{\psi}\gamma^5 \psi \qquad (2.16)$$

(where $\gamma^{\mu\nu} = \tfrac{1}{2}\left(\gamma^\mu \gamma^\nu - \gamma^\nu \gamma^\mu\right)$), namely

$$\alpha = \psi^+\psi \, (= V^0), \quad \beta = \psi^+\gamma^0\psi, \quad \chi = i\psi^+\gamma^0\gamma^5\psi, \quad \delta = \psi^+\gamma^5\psi \, (= S^0). \tag{2.17}$$

These functions are independent in the sense that, for arbitrary $\psi$, there is no functional relationship between them (this is a partial tensorial restatement of the independence of the Dirac spinor components). All other rotational scalars constructible from a Dirac spinor are reducible to functions of this set, as a result of identities holding among the bilinear covariants [10,11]. Thus, the bivector $B^{\mu\nu}$ can be expressed purely in terms of the other covariants, and the rotational scalars implied by the 4-vectors obey the identities

$$V^\mu V_\mu = -S^\mu S_\mu = \beta^2 + \chi^2, \quad V^0 S^0 = -V^i S_i = \alpha\delta, \quad (V^i)^2 = \alpha^2 - \beta^2 - \chi^2, \quad (S^i)^2 = \delta^2 + \beta^2 + \chi^2. \tag{2.18}$$

We have then that $j^0 = j^0(\alpha, \beta, \chi, \delta)$. Inserting this expression in relation (2.12) and using the relation $\gamma^0\gamma^5 = -\gamma^5\gamma^0$ we get

$$\frac{\partial j^0}{\partial \chi}\delta - \frac{\partial j^0}{\partial \delta}\chi = 0. \tag{2.19}$$

$j^0$ therefore has the following functional dependence:

$$j^0 = j^0(\alpha, \beta, \delta^2 + \chi^2). \tag{2.20}$$

The second transformation property of $j^0$ we need is that it is the zeroth component of a 4-vector. The most general form it can have as an algebraic function of the variables (2.17) is

$$j^0 = X\alpha + Y\delta + h^0 \tag{2.21}$$

where $X$ and $Y$ are Lorentz scalars and $h^0$ is a constant. Since $X$ and $Y$ can be functions only of Lorentz scalars, namely $\beta$ and $\chi$, (2.21) becomes

$$j^0 = X(\beta, \chi)\alpha + Y(\beta, \chi)\delta + h^0. \tag{2.22}$$

Comparing the linear dependence of (2.22) on $\delta$ with the general result (2.20), these expressions will be compatible only if $Y = 0$ and $X$ is independent of $\chi$. Therefore

$$j^0 = X(\beta)\alpha + h^0. \tag{2.23}$$

To complete the determination of $j^0$, we substitute (2.23) in (2.13) which reduces to

$$2\alpha \frac{\partial X}{\partial \beta}\psi^+\gamma^i\psi = 0. \tag{2.24}$$

This can be satisfied for all $\psi$ and $\psi^+$ only if $\partial X/\partial \beta = 0$ and hence we find for (2.23)

$$j^0 = g\alpha + h^0 = g\psi^+\psi + h^0, \quad g, h^0 = \text{constants} \tag{2.25}$$

To finish the story, it is a simple matter to deduce $j^i$ by substituting (2.25) in (2.14) and (2.15). We have

$$\frac{\partial j^i}{\partial \psi^b} = g\psi^{*a}\left(\gamma^0\gamma^i\right)^a_{\ b}, \quad \frac{\partial j^i}{\partial \psi^{*c}} = g\left(\gamma^0\gamma^i\right)^c_{\ a}\psi^a \tag{2.26}$$

which readily integrates to give

$$j^i = g\psi^{*a}\left(\gamma^0\gamma^i\right)^a_{\ b}\psi^b + h^i \tag{2.27}$$

where $h^i$ is constant. Combining (2.25) and (2.27) we finally deduce for the 4-current the expression

$$j^\mu = g\bar{\psi}\gamma^\mu\psi + h^\mu, \ g, h^\mu = \text{constants.} \tag{2.28}$$

Up to multiplicative and additive constants, this is of course the conventional expression given in (1.4). It is readily checked that (2.28) indeed obeys the relations (2.12)-(2.15). We have shown that this is the only conserved 4-vector implied by the Dirac equation that is a function of just the state variables.

The determination of the constants depends on the physical interpretation of the current (the only place where this need be invoked). To fix $h^\mu$ we make the physically reasonable assumption that $j^\mu = 0$ when $\psi = 0$. Then $h^\mu = 0$. If we require $j^\mu$ real then $g$ is real. If this is a probability current then $g = c$ and we recover the vector (1.4); if an electric current, $g = ec$.

It will be observed that the only properties of $j^\mu$ we needed to employ in this derivation are that it is (a) implied by the Dirac equation, (b) a function of the state, (c) conserved, and (d) has the property that $j^0$ is a rotational scalar and the zeroth component of a 4-vector. We did not need to assume any of its other key properties, which therefore may be regarded as deductions. These include that $j^\mu/g$ is bilinear, real, single-valued, gauge invariant, causal ($g^{-2}j^\mu j_\mu \geq 0$), future-pointing ($j^0/g \geq 0$), proper ($j^\mu \to \left(j^0, -j^i\right)$ under a parity transformation), and independent of the electromagnetic potentials.

## 2.3 Proof for the free Dirac equation

The above proof made essential use of equation (2.13), which follows since $A_\mu$ is an arbitrary function. This appears to make the result depend on the way in which the Dirac particle couples with an external field. To show that the result is independent of this coupling, we derive the current starting from the free Dirac equation ((2.1) with $A_\mu = 0$). Equation (2.13) is now absent and has to be replaced by an alternative equation in order to find $j^0$.

This extra relation may be obtained by eliminating $j^i$ from (2.14) and (2.15). Differentiating (2.14) with respect to $\psi^{*c}$ and (2.15) with respect to $\psi^b$ we get

$$\frac{\partial^2 j^i}{\partial \psi^{*c}\partial \psi^b} = \frac{\partial^2 j^0}{\partial \psi^{*c}\partial \psi^a}\left(\gamma^0\gamma^i\right)^a_{\ b} = \left(\gamma^0\gamma^i\right)^c_{\ a}\frac{\partial^2 j^0}{\partial \psi^{*a}\partial \psi^b}. \tag{2.29}$$

Hence

$$\Phi^c{}_a (\gamma^0 \gamma^i)^a{}_b = (\gamma^0 \gamma^i)^c{}_a \Phi^a{}_b \qquad (2.30)$$

where

$$\Phi^a{}_b = \frac{\partial^2 j^0}{\partial \psi^{*a} \partial \psi^b}. \qquad (2.31)$$

The relation (2.30) states that the matrix $\Phi$ commutes with $\gamma^0 \gamma^i$ for each $i = 1,2,3$. To use this relation to fix $\Phi$, we recall that a spinor matrix can be expanded in terms of the 16 linearly independent matrices $\gamma^Q$ comprising the unit matrix and products of the $\gamma^\mu$ s:

$$\Phi^a{}_b = d_Q \gamma^{Q\,a}{}_b. \qquad (2.32)$$

The only terms in this sum that commute with $\gamma^0 \gamma^i$ are the unit and $\gamma^5$ matrices. Thus

$$\Phi^a{}_b = d \delta^a{}_b + d_5 \gamma^{5\,a}{}_b \qquad (2.33)$$

where $d$ and $d_5$ may depend on $\psi$ and $\psi^+$.

Using the fact that, from (2.22), $j^0$ is a particular function of $\alpha$ and $\beta$, viz. $j^0 = X(\beta)\alpha + h^0$, we know that the matrix $\Phi$ must have the following form, obtained by differentiating $j^0$ twice (using (2.17)):

$$\Phi^a{}_b = \delta^a{}_b X + \alpha (\gamma^0 \psi)^a (\psi^+ \gamma^0)_b \frac{\partial^2 X}{\partial \beta^2} + \left( \alpha \gamma^{0\,a}{}_b + (\gamma^0 \psi)^a \psi^{*b} + \psi^a (\psi^+ \gamma^0)_b \right) \frac{\partial X}{\partial \beta}. \qquad (2.34)$$

We can fix the functional dependence of $j^0$ completely by comparing (2.34) with (2.33). This we do by evaluating these expressions for two sets of values of $a$ and $b$ (employing the representation (2.4)), and using the property that we can always rotate the axes to find a frame in which two components of the spinor are nonzero. When $a = 1$, $b = 4$ we obtain

$$\alpha \psi^1 \psi^{*4} \frac{\partial^2 X}{\partial \beta^2} = 0. \qquad (2.35)$$

Choosing a frame in which $\psi^1$ and $\psi^{*4}$ are nonzero gives

$$\frac{\partial^2 X}{\partial \beta^2} = 0. \qquad (2.36)$$

Since the quantities involved are scalars, this result is true independent of the frame. Next, when $a = 1$, $b = 2$ and using (2.36) we find

$$\psi^1 \psi^{*2} \frac{\partial X}{\partial \beta} = 0. \tag{2.37}$$

Choosing the frame appropriately, this gives

$$\frac{\partial X}{\partial \beta} = 0 \tag{2.38}$$

which again is a covariant result. Hence, $j^0$ is independent of $\beta$ and we obtain the result (2.25) (in (2.33), $d = g$ and $d_5 = 0$).

We can in fact generalize the technique just described to derive $j^0$ purely from (2.14) and (2.15) (i.e., without employing (2.12)), by computing the matrix (2.31) for the general expression (2.22) and comparing the result with (2.33).

### 2.4 Other conserved vectors

The theorem just proved pertains to the 4-vector that is an algebraic function of the state variables, and that is conserved by virtue of the wave equation. It does not exclude the existence of further conserved 4-vectors depending just on quantities appearing in the wave equation, i.e., $\psi$ and $\gamma^\mu$. Since the Dirac equation exhausts the constraints that apply to the wavefunction, these additional conserved vectors must lie within one of two classes: either they are conserved identically, i.e., for all $\psi$ independently of the wave equation, or they are conserved as a consequence of the wave equation. We have already shown that there is only one conserved vector depending just on the state variables so, if they exist, additional vectors lying in the second class must depend on the derivatives of the fields. The same will be true for the vectors in the first class. Although the spinor functions for which a vector is identically conserved are arbitrary (subject to differentiability etc.), we are interested only in those which obey Dirac's equation. It is known that there exist non-trivial identities connecting the fields and their first derivatives [10]. However, we have shown that, if the spinor function obeys the Dirac equation, the only associated conserved 4-vector that is just an algebraic function of $\psi$ and $\psi^+$ is $j^\mu$. Hence, an identically divergence-free 4-vector must depend on the derivatives of the fields. We now give some examples and examine the general features of the additional vectors.

Denoting the set of additional vectors by $a_n^\mu$, the total conserved current associated with the Dirac equation will be

$$\bar{j}^\mu = j^\mu + \sum_n a_n^\mu, \quad \partial_\mu \bar{j}^\mu = 0. \tag{2.39}$$

Any 4-vector $a^\mu$ with zero divergence can be written as a 4-curl:

$$a^\mu = \partial_\nu \xi^{\mu\nu}, \quad \partial_\mu a^\mu = 0, \tag{2.40}$$

where $\xi^{\mu\nu} = -\xi^{\nu\mu}$. We shall henceforth restrict this notation to identify identically conserved vectors. In the case that the tensor field $\xi^{\mu\nu}$ is a local function of the fields and their derivatives it will not contribute to the conserved quantity (2.6). We have

$$\int \bar{j}^0 \, d^3x = \int j^0 \, d^3x + \int \partial_i \xi^{0i} d^3x. \tag{2.41}$$

Applying the usual boundary conditions on the wavefunction, the second term vanishes if, e.g., $\xi^{0i}$ is an algebraic function of $\psi$. An example of an identically conserved vector of this type, built from the bivector $B^{\mu\nu}$ (cf. (2.16)), is

$$a^\mu = k\partial_\nu(\bar{\psi}\gamma^{\mu\nu}\psi), \quad \partial_\mu a^\mu = 0, \quad k = \text{constant}, \tag{2.42}$$

which is conserved due to the antisymmetry in the $\gamma$ matrices.

The Dirac current can be usefully decomposed into terms that involve identically conserved vectors. A well known example is the Gordon decomposition (in which (2.42) appears) which follows by replacing the fields in the Dirac current (1.4) by their derivatives, using the field equations (2.1) and (2.2):

$$J_\mu = \frac{i\hbar}{2m}\left(\bar{\psi}\partial_\mu\psi - (\partial_\mu\bar{\psi})\psi\right) - \frac{e}{m}\bar{\psi}\psi A_\mu + \frac{i\hbar}{2m}\partial^\nu(\bar{\psi}\gamma_{\mu\nu}\psi). \tag{2.43}$$

This example is important for another reason, for it brings us to the second class of additional vectors, those that are conserved as a consequence of the wave equation. The first two terms in (2.43) taken together form a proper, gauge invariant 4-vector:

$$G_\mu = \frac{i\hbar}{2m}\left(\bar{\psi}\partial_\mu\psi - (\partial_\mu\bar{\psi})\psi\right) - \frac{e}{m}\bar{\psi}\psi A_\mu. \tag{2.44}$$

Since the third term in (2.43) is conserved identically, $G^\mu$ is conserved by virtue of the Dirac equation (as may be readily checked). As expected, (2.44) depends on the field derivatives. (Another way of arriving at this vector is to observe that a Dirac spinor obeys the Klein-Gordon equation and hence will appear in the conserved current implied by the latter (see §4); the first two terms in (2.44) are just the spinor version of (4.19).) Indeed, we may derive an infinite hierarchy of conserved vectors from the Dirac current, involving ever higher orders of derivatives of the fields, by extending the technique implicit in the Gordon decomposition. Taking the free case ($A_\mu = 0$) for simplicity, we can, for example, substitute for the fields in (2.44) from the wave equations (2.1) and (2.2) to obtain

$$G_\mu = \frac{\hbar^2}{2m^2c}\left((\partial_\nu\bar{\psi})\gamma^\nu\partial_\mu\psi + (\partial_\mu\bar{\psi})\gamma^\nu\partial_\nu\psi\right). \tag{2.45}$$

This expression can be rearranged to give

$$G^\mu = \frac{\hbar^2}{2m^2c}\left[-\bar{\psi}\gamma^\nu\partial_\nu\partial^\mu\psi - (\partial_\nu\partial^\mu\bar{\psi})\gamma^\nu\psi + \partial_\nu(\bar{\psi}\gamma^\mu\partial^\nu\psi + (\partial^\nu\bar{\psi})\gamma^\mu\psi)\right]$$
$$+ \frac{\hbar^2}{2m^2c}\partial_\nu\left(\bar{\psi}\gamma^\nu\partial^\mu\psi - \bar{\psi}\gamma^\mu\partial^\nu\psi + (\partial^\mu\bar{\psi})\gamma^\nu\psi - (\partial^\nu\bar{\psi})\gamma^\mu\psi\right). \tag{2.46}$$

The second term is the divergence of an antisymmetric tensor and is conserved identically. The first term is therefore a (second-order) vector that is conserved as a consequence of the Dirac equation.

It is clear that all higher-order vectors $b^\mu$, $b'^\mu$ constructed in this way differ from one another by an identically conserved 4-vector:

$$b^\mu = b'^\mu + \partial_\nu \xi^{\mu\nu} \tag{2.47}$$

In particular, they are all related to the Dirac current by the equation

$$b^\mu = J^\mu + \partial_\nu \xi^{\mu\nu} \tag{2.48}$$

where $\partial_\nu \xi^{\mu\nu}$ depends on the same order of derivatives as $b^\mu$. Since the identically conserved vectors do not contribute to the global conserved quantity $P$, each of the Dirac-conserved vectors implies the same global value as the Dirac current:

$$P = \int b^0 \, d^3x = \int J^0 \, d^3x \tag{2.49}$$

This may be checked by using the wave equation to reduce the integrand in the second member of (2.49) to $c\psi^+\psi$. Thus, for the vector (2.44) we have

$$G^0 = c\psi^+\psi - \frac{i\hbar}{2m} \partial_i \left( \psi^+ \gamma^i \psi \right) \tag{2.50}$$

and integrating over all space the second term vanishes.

The global quantity $P$ is therefore compatible with an infinite variety of choices for the local current. This does not undermine the argument of §2.1 that $P$ is an algebraic functional since, as we have just seen, the global values of the zeroth components of all the currents are reducible to the integral of an algebraic function. It does, however, bring out clearly that the requirement of §2.1 that the local current is also just an algebraic function is an additional assumption. It follows from the requirement that the current be a covariant function of the fields. Although, as noted in §2.1, we can always use the wave equation to eliminate the time derivatives of the fields in the vectors $b^\mu$, these will depend on the spatial derivatives (an example is given in (2.50)) and hence they are not functions just of the state.

One might argue that, of all the vectors, the Dirac current is preferred in that it is timelike and has a non-negative zeroth component for all wavefunctions. This, however, is an accident of the Dirac algebra; no such argument could be applied in the Klein-Gordon case, for example, where likewise a plethora of conserved currents occur (§4.2). In general, we cannot rule out a key role for the identically-conserved vectors (cf. the Pauli current in §5 where the last term in (5.3) has vanishing divergence identically). There does not appear to be any naturally occurring constraint in the theory of a Dirac particle subject to external forces that would limit the options. We must therefore regard the choice of vector $b^\mu$ as a component of the unique specification of the state of the system, additional to the fields.

Apart from conserved vectors depending only on quantities appearing in the wave equation, there exists one further class of conserved vectors: those that depend on quantities not contained in the wave equation. In that case, it is possible that there might exist further conserved vectors that are algebraic functions of $\psi$, if suitable additional functions are introduced (for an example of this sort see §4.2).

**3 Coupled Maxwell-Dirac equations**

In the demonstration given in §2.2 it was assumed that the electromagnetic potentials $A_\mu$ appearing in the Dirac equation (2.1) are prescribed functions. We now show how to extend the uniqueness proof to include the case where the particle and the electromagnetic field mutually interact, and are subject to the Maxwell-Dirac equations:

$$\gamma^\mu(i\hbar\partial_\mu - eA_\mu)\psi = mc\psi, \quad \partial_\nu F^{\mu\nu} = -j^\mu \tag{3.1}$$

where $F_{\mu\nu} = \partial_\mu A_\nu - \partial_\nu A_\mu$. Substituting the latter expression into Maxwell's equations gives

$$\partial_\nu \partial^\nu A^\mu - \partial_\nu \partial^\mu A^\nu = j^\mu. \tag{3.2}$$

In the Lorenz gauge ($\partial_\nu A^\nu = 0$) (3.2) reduces to the inhomogeneous wave equation and has the solution [1]

$$A^\mu(x) = \mathring{A}^\mu(x) + \int G(x - x') j^\mu(x') d^4 x' \tag{3.3}$$

where $G(x - x')$ is the Green function and $\mathring{A}^\mu(x)$ obeys the homogeneous wave equation. The $A_\mu$ appearing in Dirac's equation is now shorthand for (3.3).

In this context we may repeat the argument of §2.1 that excludes dependence of $j^\mu$ on the derivatives of $\psi$ and $\psi^+$. Since $A_\mu$ is now a function of the fields and $\mathring{A}_\mu$, the current will have the following dependence: $j^\mu = j^\mu(\psi, \psi^+, \mathring{A})$. We aim to determine this dependence; in the usual Maxwell-Dirac equations $j^\mu = eJ^\mu$, of course.

The conservation equation (2.5) here takes the form (2.7) with $A_\mu$ replaced by $\mathring{A}_\mu$. As in §2.2, we may argue that, since the derivatives of $\mathring{A}_\mu$ are arbitrary, $j^\mu$ is independent of $\mathring{A}_\mu$. The conservation equation therefore reduces to (2.7) and the argument to derive $j^\mu$ proceeds as follows. Equation (2.7) is once again written in the form (2.11). From this we use the independence of the field derivatives to derive equations (2.14) and (2.15). Using the condition that $\mathring{A}_\mu$ is an arbitrary function we can in addition deduce equations (2.12) and (2.13). We want, however, to include the possibility that $\mathring{A}_\mu = 0$. In that case we cannot deduce the separate relations (2.12) and (2.13) (because $A_\mu$ depends on the fields). Instead, we have the single equation

$$\frac{\partial j^0}{\partial \psi}(e\gamma^0 \gamma^\mu A_\mu \psi + mc\gamma^0 \psi) - (e\psi^+ \gamma^0 \gamma^\mu A_\mu + mc\psi^+ \gamma^0)\frac{\partial j^0}{\partial \psi^+} = 0. \tag{3.4}$$

Inserting in this relation the general functional form $j^0 = j^0(\alpha, \beta, \chi, \delta)$ gives

$$\bar\psi\left(\frac{\partial j^0}{\partial \beta}\gamma^0\gamma^i + i\frac{\partial j^0}{\partial \chi}\gamma^5\gamma^0\gamma^i\right)\psi A_i + imc\left(\frac{\partial j^0}{\partial \delta}\chi - \frac{\partial j^0}{\partial \chi}\delta\right) = 0. \tag{3.5}$$

Since from (3.3) $A_i$ is independent of $j^0$ the second term in (3.5) must vanish. We thus obtain equation (2.19) and hence the expression (2.23) for $j^0$. Inserting this in (3.5) then gives

$$\alpha \frac{\partial X}{\partial \beta} \bar{\psi} \gamma^0 \gamma^i \psi A_i = 0 \tag{3.6}$$

which can be true for all fields only if $\partial X/\partial \beta = 0$. We thus obtain (2.25) and the derivation continues as in §2.2, culminating in (2.28). Although this derivation relies on a particular choice of gauge, the final result (2.28) is gauge invariant and we will therefore obtain the same expression whatever gauge is used.

We conclude that, if the current depends just on the state variables, the conventional expression for the Dirac current is the only one that can be introduced as a source of the electromagnetic field in the case of coupled systems (with $g = ec$).

**4 The Klein-Gordon current**

**4.1 Derivation of the unique current**

We shall derive here the unique conserved current implied by the free Klein-Gordon equation and the complex conjugate equation,

$$\partial^\mu \partial_\mu \psi + \left(m^2 c^2/\hbar^2\right)\psi = 0, \qquad \partial^\mu \partial_\mu \psi^* + \left(m^2 c^2/\hbar^2\right)\psi^* = 0, \tag{4.1}$$

assuming that the current is a function only of the state variables. Here $\psi$ is a Lorentz scalar field and $\psi$ and $\psi^*$ are regarded as independent functions.

Applying the argument of §2.1, the second-order character of the wave equation implies that the current will be a function of the fields and their first derivatives (the state variables): $j^\mu = j^\mu\left(\psi, \psi^*, \partial_\nu \psi, \partial_\nu \psi^*\right)$. The conservation equation (2.5) therefore becomes

$$\frac{\partial j^\mu}{\partial \psi} \partial_\mu \psi + \frac{\partial j^\mu}{\partial \psi^*} \partial_\mu \psi^* + \frac{\partial j^\mu}{\partial\left(\partial_\nu \psi\right)} \partial_\mu \partial_\nu \psi + \frac{\partial j^\mu}{\partial\left(\partial_\nu \psi^*\right)} \partial_\mu \partial_\nu \psi^* = 0. \tag{4.2}$$

If we wanted to extend the treatment to include the electromagnetic potentials in the wave equation, an additional term corresponding to that in (2.7) would have to be added to the left-hand side of (4.2). Unlike the Dirac case, this term will be finite since derivatives of the potentials appear in the wave equation (and consequently the potentials appear in the current).

To obtain an identity, we substitute for the second-order time derivatives from (4.1):

$$\frac{\partial j^\mu}{\partial \psi}\partial_\mu\psi + \frac{\partial j^\mu}{\partial \psi^*}\partial_\mu\psi^* + \frac{\partial j^0}{\partial(\partial_0\psi)}\left(\partial_i\partial_i\psi - \left(m^2c^2/\hbar^2\right)\psi\right) + \left(\frac{\partial j^0}{\partial(\partial_i\psi)} + \frac{\partial j^i}{\partial(\partial_0\psi)}\right)\partial_0\partial_i\psi + \frac{\partial j^i}{\partial(\partial_k\psi)}\partial_i\partial_k\psi$$

$$+ \frac{\partial j^0}{\partial(\partial_0\psi^*)}\left(\partial_i\partial_i\psi^* - \left(m^2c^2/\hbar^2\right)\psi^*\right) + \left(\frac{\partial j^0}{\partial(\partial_i\psi^*)} + \frac{\partial j^i}{\partial(\partial_0\psi^*)}\right)\partial_0\partial_i\psi^* + \frac{\partial j^i}{\partial(\partial_k\psi^*)}\partial_i\partial_k\psi^* = 0.$$

(4.3)

Since $j^\mu$ is independent of the second derivatives of the fields, the coefficient of each of these terms is zero and we get the following relations:

$$\frac{\partial j^\mu}{\partial \psi}\partial_\mu\psi + \frac{\partial j^\mu}{\partial \psi^*}\partial_\mu\psi^* - \left(m^2c^2/\hbar^2\right)\left(\frac{\partial j^0}{\partial(\partial_0\psi)}\psi + \frac{\partial j^0}{\partial(\partial_0\psi^*)}\psi^*\right) = 0 \qquad (4.4)$$

$$\frac{\partial j^0}{\partial(\partial_0\psi)} = -\frac{\partial j^1}{\partial(\partial_1\psi)} = -\frac{\partial j^2}{\partial(\partial_2\psi)} = -\frac{\partial j^3}{\partial(\partial_3\psi)}, \quad \frac{\partial j^0}{\partial(\partial_0\psi^*)} = -\frac{\partial j^1}{\partial(\partial_1\psi^*)} = -\frac{\partial j^2}{\partial(\partial_2\psi^*)} = -\frac{\partial j^3}{\partial(\partial_3\psi^*)} \qquad (4.5)$$

$$\frac{\partial j^i}{\partial(\partial_k\psi)} = \frac{\partial j^i}{\partial(\partial_k\psi^*)} = 0, \; i \neq k \qquad (4.6)$$

$$\frac{\partial j^0}{\partial(\partial_i\psi)} + \frac{\partial j^i}{\partial(\partial_0\psi)} = \frac{\partial j^0}{\partial(\partial_i\psi^*)} + \frac{\partial j^i}{\partial(\partial_0\psi^*)} = 0, \quad i = 1, 2, 3. \qquad (4.7)$$

To use these relations to fix the current, we note that the most general 4-vector one can construct from a complex scalar field and its first derivatives is

$$j_\mu = X\partial_\mu\psi + Y\partial_\mu\psi^* + h_\mu \qquad (4.8)$$

where the Lorentz scalars $X$ and $Y$ depend on the fields $\psi$, $\psi^*$ and their first derivatives, and $h_\mu$ is constant. The scalar character of $X$ and $Y$ means they can depend on the derivatives only through the independent functions

$$\partial^\mu\psi\partial_\mu\psi, \quad \partial^\mu\psi^*\partial_\mu\psi^*, \quad \partial^\mu\psi\partial_\mu\psi^*. \qquad (4.9)$$

Inserting the expression (4.8) in the first member of (4.6) we obtain

$$\frac{\partial X}{\partial(\partial_k\psi)}\partial^i\psi + \frac{\partial Y}{\partial(\partial_k\psi)}\partial^i\psi^* = 0, \; i \neq k. \qquad (4.10)$$

This relation and the second member of (4.6) may be used to determine the dependence of $X$ and $Y$ on the field derivatives. This we do by using the property of a complex 3-vector (here $\partial^i\psi$) that, by means of a suitable spatial rotation, we can always find a frame in which the real or imaginary part of one component is zero. Consider the component $\partial^1\psi$ and choose the frame so that it is real. Then (4.10) gives

$$\left(\frac{\partial X}{\partial(\partial_k\psi)} + \frac{\partial Y}{\partial(\partial_k\psi)}\right)\partial^1\psi = 0, \ k = 2,3. \tag{4.11}$$

Hence, in this frame

$$\frac{\partial}{\partial(\partial_k\psi)}(X + Y) = 0, \ k = 2,3 \tag{4.12}$$

and $X + Y$ is independent of $\partial_k\psi$, $k = 2,3$. Now choose a frame in which $\partial^1\psi$ is pure imaginary. Then

$$\left(\frac{\partial X}{\partial(\partial_k\psi)} - \frac{\partial Y}{\partial(\partial_k\psi)}\right)\partial^1\psi = 0, \ k = 2,3 \tag{4.13}$$

and so

$$\frac{\partial}{\partial(\partial_k\psi)}(X - Y) = 0, \ k = 2,3. \tag{4.14}$$

Thus, in this frame, $X - Y$ is independent of $\partial_k\psi$, $k = 2,3$. But $X$ and $Y$ are scalars so the values of $X \pm Y$ are independent of the frame. That is, both (4.12) and (4.14) are true and we conclude that $X$ and $Y$ are each independent of $\partial_k\psi$, $k = 2,3$. Relativistic covariance (cf. (4.9)) then implies they are each independent of $\partial_3\psi$ and $\partial_0\psi$. Starting from the second relation in (4.6), a similar argument shows that these functions are also independent of $\partial_\mu\psi^*$ for all $\mu$. These results are consistent with the relations (4.5) (note: $\partial_i\psi = -\partial^i\psi$) and (4.7).

Hence, $X$ and $Y$ are functions of just $\psi$ and $\psi^*$. To find this dependence, we substitute (4.8) in (4.4) to get

$$\frac{\partial X}{\partial \psi}\partial^\mu\psi\partial_\mu\psi + \frac{\partial Y}{\partial \psi^*}\partial^\mu\psi^*\partial_\mu\psi^* + \left(\frac{\partial Y}{\partial \psi} + \frac{\partial X}{\partial \psi^*}\right)\partial^\mu\psi\partial_\mu\psi^* - \left(m^2c^2/\hbar^2\right)\left(X\psi + Y\psi^*\right) = 0. \tag{4.15}$$

Since in this equation the coefficients of the field derivatives and the final term are independent of the derivatives, and the latter appear just in terms of the independent scalars (4.9), we must have

$$\frac{\partial X}{\partial \psi} = \frac{\partial Y}{\partial \psi^*} = \frac{\partial Y}{\partial \psi} + \frac{\partial X}{\partial \psi^*} = 0, \tag{4.16}$$

$$\frac{X}{\psi^*} = -\frac{Y}{\psi}. \tag{4.17}$$

The first two relations in (4.16) imply that $X = f_1(\psi^*) - d_1$, $Y = f_2(\psi) - d_2$ where $d_1$ and $d_2$ are constants. These results then imply that each side in (4.17) equals a constant, say $g$, and we obtain

$$X = g\psi^* + d_1, \ Y = -g\psi + d_2 \tag{4.18}$$

(which obeys the last relation in (4.16)). These expressions satisfy (4.17) only if $d_1 = d_2 = 0$. Substituting in (4.8) therefore gives finally for the current

$$j_\mu = g(\psi^* \partial_\mu \psi - \psi \partial_\mu \psi^*) + h_\mu, \quad g, h_\mu = \text{constants}. \tag{4.19}$$

Assuming that $j^\mu = 0$ when $\psi = 0$ we have $h^\mu = 0$. Choosing the constant $g = i\hbar/2m$, we obtain the usual Klein-Gordon current. Apart from the choice of constants, this is the only conserved 4-vector implied by the Klein-Gordon wave equation that depends just on the fields and their first derivatives.

**4.2 Other conserved vectors**

As in the Dirac case (§2.4), this derivation does not exclude the existence of other 4-vectors $a^\mu$, depending just on the wavefunction and its derivatives, which are conserved either identically (of the form (2.40), i.e., $a^\mu = \partial_\nu \xi^{\mu\nu}$) or by virtue of the wave equation. In both cases our results show that such vectors must depend on at least the second derivatives of the fields. An example of an identically conserved vector is

$$a_\mu = k \partial^\nu (\partial_\mu \psi \partial_\nu \psi^* - \partial_\nu \psi \partial_\mu \psi^*), \quad k = \text{constant}, \tag{4.20}$$

which is divergenceless due to the antisymmetric combination of derivatives. This vector appears in the analogue for the Klein-Gordon current of the Gordon decomposition of the Dirac current, (2.43). Substituting for the fields in (4.19) from (4.1) and rearranging we get

$$j_\mu = \frac{i\hbar^3}{2m^3 c^2} (\partial^\nu \partial_\mu \psi \partial_\nu \psi^* - \partial_\nu \psi \partial^\nu \partial_\mu \psi^*) + \frac{i\hbar^3}{2m^3 c^2} \partial^\nu (\partial_\nu \psi \partial_\mu \psi^* - \partial_\mu \psi \partial_\nu \psi^*). \tag{4.21}$$

The second term is identically conserved and hence the first term provides an example of a (second-order) vector that is conserved in virtue of the wave equation.

We can construct conserved vectors depending on just the first derivatives other than (4.19) only if we allow a dependence on quantities not contained in the wave equation. An example of this sort has been given previously [8]: if $k^\mu$ is some constant 4-vector and $T^{\mu\nu}$ is the energy-momentum tensor of the Klein-Gordon field, the 4-vector $l^\mu = k_\nu T^{\mu\nu}$ is conserved by virtue of the Klein-Gordon equation. This example is valuable since, if $k^\mu$ is future-causal, $l^\mu$ will be also.

If we desire to define for the relativistic spin 0 system a conserved 4-current having the properties of the Dirac current, *viz.* timelike with a positive zeroth component, our results show that we are obliged either to seek a vector involving at least the second derivatives of the field, or to introduce some further structure not contained in the Klein-Gordon equation. In the former case no such vector with the required properties is apparently known; in the latter, an example has just been given. These remarks conflict with a recent claim to have constructed a conserved future-timelike vector purely from the Klein-Gordon wavefunction and its first derivatives (of the form (4.8) for a particular choice of X and Y depending on the derivatives) [12].

## 5 The Pauli and Schrödinger currents

In non-relativistic physics the behaviour of a spin $\frac{1}{2}$ particle of charge $e$ and magnetic moment $\kappa$ is governed by the Pauli equation for a two-component spinor field $\phi^A(\mathbf{x},t)$, $A = 1,2$:

$$i\hbar\frac{\partial\phi}{\partial t} = \left[\frac{1}{2m}(-i\hbar\nabla - e\mathbf{A})^2 + V + \kappa\mathbf{B}.\sigma\right]\phi \qquad (5.1)$$

where $V$ is the total external scalar potential, and $\mathbf{B} = \nabla \times \mathbf{A}$. The techniques of §2 and §4 can be adapted to derive the unique current implied by this equation when it is assumed that the current depends only on the state functions. However, there is no need to do this because, when applied to electrons (for which $\kappa = -e\hbar/2m$) and other systems whose relativistic aspects are described by the Dirac equation, the Pauli equation is its non-relativistic limit (for a discussion of the subtleties of the limiting procedure see [13]). That is, the Pauli current must be whatever is found as the non-relativistic limit of the current associated with the Dirac equation. Under the assumption that the relativistic current depends only on the state, the Pauli current so obtained will be unique since we have already proved its relativistic forebear to be unique.

The limiting Pauli current $(\rho, \mathbf{j})$ implied by the Dirac current (2.42) is

$$\rho\left(\equiv J^0/c\right) = \phi^+\phi \qquad (5.2)$$

$$\mathbf{j} = (\hbar/2mi)\left(\phi^+\nabla\phi - (\nabla\phi^+)\phi\right) - (e\rho/m)\mathbf{A} + (1/m)\nabla \times (\rho\mathbf{s}) \qquad (5.3)$$

where $\mathbf{s} = (\hbar/2\rho)\phi^+\sigma\phi$ is the spin vector. This is the required unique current associated with the Pauli equation, and is of course the usual expression.

As shown elsewhere [13], the coupled Maxwell-Dirac equations do not have a non-relativistic limit (in the sense that there is no limiting case of the equations that is Galilean covariant) and hence the issue of the uniqueness of a limiting current does not arise in that case.

For the special case where the magnetic field may be neglected and the system is in a spin eigenstate,

$$\phi^A(\mathbf{x},t) = \Psi(\mathbf{x},t)\chi^A, \quad \chi^+\chi = 1, \qquad (5.4)$$

the Pauli equation reduces to the Schrödinger equation for the function $\Psi$ but the current (5.3) does not coincide with the conventional Schrödinger expression for the current given in (1.2) [14]. We have rather

$$\rho = \Psi^*\Psi, \quad \mathbf{j} = (\hbar/2mi)\left(\Psi^*\nabla\Psi - (\nabla\Psi^*)\Psi\right) + (1/m)\nabla(\Psi^*\Psi) \times \mathbf{s}, \quad \mathbf{s} = (\hbar/2)\chi^+\sigma\chi \qquad (5.5)$$

and $\mathbf{j}$ contains a generally non-vanishing, spin-dependent term.

We can apply a similar argument to obtain the unique current associated with the Schrödinger equation by regarding this as a limit of the (unique) Klein-Gordon current (4.19). As expected, the formulas obtained are those given in (1.2). It will be noted that the Schrödinger equation is common to the description of both the spin 0 theory and the spin $\frac{1}{2}$ theory (for an eigenstate), but that the form of the vector **j** differs in the two cases. The reason is that the total wavefunction in the latter case is (5.4).

## 6 Discussion

We have implemented Fock's programme in the context of quantum-mechanical currents and found that in the Dirac, Maxwell-Dirac and Klein-Gordon cases the usual expressions are the only currents that depend just on the relevant state variables. This analysis only partially answers the question that motivated this investigation for, as we have shown, there exist further conserved vectors associated with the Dirac and Klein-Gordon equations. Nevertheless, the results are useful in establishing some general characteristics of the other vectors (e.g., that they must depend on other variables such as higher field derivatives). A remaining question is why we should select preferentially as the physical vectors those depending only on the functions of state.

Further light is thrown on this matter by considering the analogous case of energy-momentum tensors, for there, for certain physical systems, we do not use expressions depending only on the state. Fock argued that dependence on just the state variables is a common property of all the tensors in common use, and for the systems to which he applied his method (the electromagnetic and hydrodynamic) this is indeed true. But it does not appear to be generally so. For the Dirac particle, where the state is exhausted by the functions $\psi$ and $\psi^+$, we would expect that all its associated local conserved quantities will be algebraic functions of the fields. Yet the conventional expression for the (canonical, unsymmetrized) Dirac energy-momentum tensor, which is proportional to

$$T^{\mu\nu} = \overline{\psi}\gamma^{\mu}\partial^{\nu}\psi - \left(\partial^{\mu}\overline{\psi}\right)\gamma^{\nu}\psi, \tag{6.1}$$

includes derivatives of the fields. For the Klein-Gordon system, although the usual energy-momentum tensor depends on just the state variables, it has been found useful (to obtain a renormalizable theory) to employ an "improved" tensor that includes an additive identically conserved tensor depending on the second derivatives of the fields [15]. And even in the case of electromagnetism, the debate on the correct form of the energy-momentum tensor goes back a century [16]. Having noted that the details of local electromagnetic field energy and momentum conservation are of fundamental importance to our understanding of matter-field interactions, Feynman [17] pointed out that this account is dependent on the precise forms we adopt for the locally conserved quantities, and that these could differ from the conventional expressions (he remarks that the other tensors will involve derivatives of the fields, which is a consequence of Fock's theorem).

As noted in §2.1, by use of the relevant wave equations we can always remove the time derivatives of the fields from the various tensors just cited to leave (non-covariant) representations of the tensor components in terms of the state variables and their spatial derivatives. For example, for the 00-component of the Dirac expression (6.1) we obtain

$$T^{00} = T_{\mu}^{\mu} - T_{i}^{i} = \left(2mc/i\hbar\right)\overline{\psi}\psi + \left(\partial_{i}\overline{\psi}\right)\gamma^{i}\psi - \overline{\psi}\gamma^{i}\partial_{i}\psi \tag{6.2}$$

(which is the sum of terms that individually do not transform as 00-components). However, the space derivatives in (6.2) cannot be removed, e.g., by symmetrization or by use of the wave equation, and they appear in the global conserved quantities. Hence, although it is true that local conserved quantities may be written so that they depend on time derivatives of order one less than appear in the relevant wave equations, generally they are not functions of just the state variables (and in a covariant representation the time derivatives must be present).

It appears therefore that, although intuitively attractive, and although it happens that the conventional expressions for quantum currents obey it, Fock's assumption of state-dependence is too stringent and cannot be generally applied as a physical selection rule. Why certain conserved quantities associated with certain physical systems do nevertheless obey Fock's principle deserves further investigation.

There are other possibilities for solving the uniqueness problem. For example, the current is in principle measurable by virtue of its effects on other systems (e.g., as a source of an electromagnetic field). And as mentioned in §1, we may invoke the principle of minimal coupling which uniquely selects the interaction term for a coupled Maxwell-Dirac system. Also, we could address the problem starting from quantum field theory. An argument for the uniqueness of the energy-momentum tensor operator has been given by Wald [18] on the basis of a set of axioms that such an object might be expected to obey (it is mentioned by Wald that his method applies also to the electric current operator).

It may seem unsatisfactory that there is no conclusive theoretical argument that secures uniqueness within the first quantized theory of particles in external fields. On the other hand, maybe the additional conserved currents that are naturally associated with the Dirac and Klein-Gordon equations beyond the conventional forms are of physical significance, depending on the context in which the theories are applied.

I would like to thank Rick Leavens who first raised with me the problem of the uniqueness of the pilot-wave guidance equation that led to the more general investigation of quantum-mechanical currents presented here, C.R. Hagen for correspondence, and Chris Philippidis and Harvey Brown for many valuable discussions.